\documentclass[aps,showpacs,twocolumn,floatfix,prl]{revtex4}
\usepackage{graphicx}
\usepackage{amsfonts}
\usepackage{amsmath}
\usepackage{amssymb}

\newcommand{\beq}{\begin{equation}}
\newcommand{\eeq}{\end{equation}}
\newcommand{\beqn}{\begin{eqnarray}}
\newcommand{\eeqn}{\end{eqnarray}}
\newcommand{\bearr}{\begin{array}}
\newcommand{\enarr}{\end{array}}

\begin{document}


\title{Quantum resonances and rectification of driven cold atoms
in optical lattices}

\author{S. Denisov, L. Morales-Molina, and S. Flach}
\affiliation{Max-Planck-Institut f\"ur Physik Komplexer
Systeme, N\"othnitzer Str. 38, 01187 Dresden, Germany}

\date{\today}
\vskip 2.cm
\begin{abstract}

Classical Hamiltonian ratchets have been
recently successfully realized using cold atoms in driven optical
lattices.
Here we study the current rectification of the motion of a quantum particle
in a periodic potential exposed to an external ac field.
The dc current appears due to the desymmetrization of Floquet
eigenstates, which become transporting.
Quantum dynamics enhances the dependence of the current on
the initial phase of the driving field.
By changing the laser field parameters which control
the degree of space-time asymmetry, Floquet eigenstates
are tuned through avoided crossings.
These quantum resonances induce
resonant changes of the resulting
current. The width, strength and position of these quantum resonances
are tunable using control parameters of the experimental
realization with cold atoms.
\end{abstract}

\maketitle

The \textit{ratchet} effect, i.e. the possibility to rectify
transport with the help of zero-mean perturbations, has been discussed
in order to explain mechanisms of
microbiological motility \cite{first}, and was applied to other
situations as well \cite{Reim}, including quantum systems \cite{Qu1}.
The dissipationless limit of Hamiltonian ratchets \cite{Flach1, Flach2, Den, Ketz}
and the corresponding symmetry predictions \cite{Flach1} have been
recently successfully studied with cold Rubidium and Cesium atoms in optical
lattices with a two-harmonics driving and a tunable weak dissipation \cite{ren1}.
In these experiments, the mechanism of the Sisyphus cooling
\cite{Sis} has been used in order to furnish initial conditions in
form of an \textit{optical lattice}: an ensemble of atoms localized
in the  wells of a periodic potential. In the momentum space this
corresponds to a narrow distribution near the momentum $p=0$. This is essential
for the observation of the rectification effect, since for broad
initial distributions the asymptotic current tends to zero.

The dissipationless case can
be readily implemented in atom optics by using laser beams which generate far detuned standing
waves \cite{rew}. Appropriate time-dependent forces can be applied to
the atoms by phase modulating the lattice beams \cite{ren1}.
The reachable quantum regime of cold atoms becoming coherent matter waves
calls for a study of a quantum Hamiltonian ratchet.
Previous studies of quantum ratchets were based on the kicked rotor
model \cite{Ketz,Qu1,Qu2}, which is easily treated numerically,
but posesses a
broad frequency spectrum of the kick drive.
The above mentioned experimental realization of
a two-frequency driven classical Hamiltonian ratchet instead suggests to consider
the corresponding quantum problem of a
particle moving in a spatially periodic potential under the influence of a
two frequency  ac force of zero mean and to search for
quantum peculiarities of the
current rectification.

The Hamiltonian for a particle with position $x$ and momentum $p$ is \cite{ren1}
\begin{equation}
H(x,p,t)=\frac{p^{2}}{2}+(1+\cos(x))-x E(t),
\label{eq:ham}
\end{equation}
where $E(t)$ is an external periodic field of zero mean,
$E(t+T)=E(t)$, $\int_0^T E(t) dt=0$.
For the classical case
there are two symmetries which need to be broken
to fulfill the necessary conditions for a nonzero dc current \cite{Flach1}. If $E(t)=-E(t+T/2)$
is shift symmetric, then (\ref{eq:ham}) is invariant under symmetry
\begin{equation}
S_{a}: (x, p, t) \rightarrow (-x, -p, t+T/2). \label{eq:Sa}
\end{equation}
If $E(t)=E(-t)$ is symmetric, then (\ref{eq:ham}) is invariant under
\begin{equation}
S_{b}: (x, p, t) \rightarrow (x, -p, -t).  \label{eq:Sb}
\end{equation}
The phase space of system ({\ref{eq:ham}) is mixed, containing both regular regions
and a chaotic layer around $p=0$.
Whenever the system
(\ref{eq:ham}) possesses any of the two symmetries $S_{a}$ and/or $S_{b}$,
directed transport is forbidden inside the chaotic layer \cite{Flach1}.

The two frequency driving
\begin{equation}
E(t, t_{0})=E_{1}\cos (\omega (t-t_{0}))+E_{2}\cos (2\omega
(t-t_{0})+\theta) \label{eq:driv}
\end{equation}
ensures that  for $E_1,E_{2} \neq 0$ $S_a$ is always violated. In
addition $S_b$ is violated for $\theta \neq 0,\pm \pi$. The
appearance of a nonzero dc current $J_{ch}=\lim_{t\rightarrow
\infty} 1/t \int_{t_0}^t p(t') dt'$ in this case is due to a
desymmetrization of the chaotic layer structure
(Fig.\ref{Fig:husi}a). It induces a desymmetrization of the events
of directed motion to the right and left \cite{D&F}. Due to
ergodicity inside the layer, the asymptotic current is independent
of the initial phase $t_{0}$, for initial conditions located inside
the chaotic layer. With the specific choice of the driving
(\ref{eq:driv}) it follows $J_{ch}(\theta)=-J_{ch}(-\theta)$ and
$J_{ch}(\theta)=-J_{ch}(\theta+\pi)$ \cite{Flach1}. From
perturbation theory it follows $J_{ch}\sim E_{1}^2 E_{2}\sin \theta$
\cite{Flach1, Flach2}. An efficient sum rule allows to compute the
average current $J_{ch}$ by proper integration over the chaotic
layer \cite{Ketz}.

\begin{center}
\begin{figure}
\begin{tabular}{cc}
\includegraphics[width=3.5cm,height=4cm]{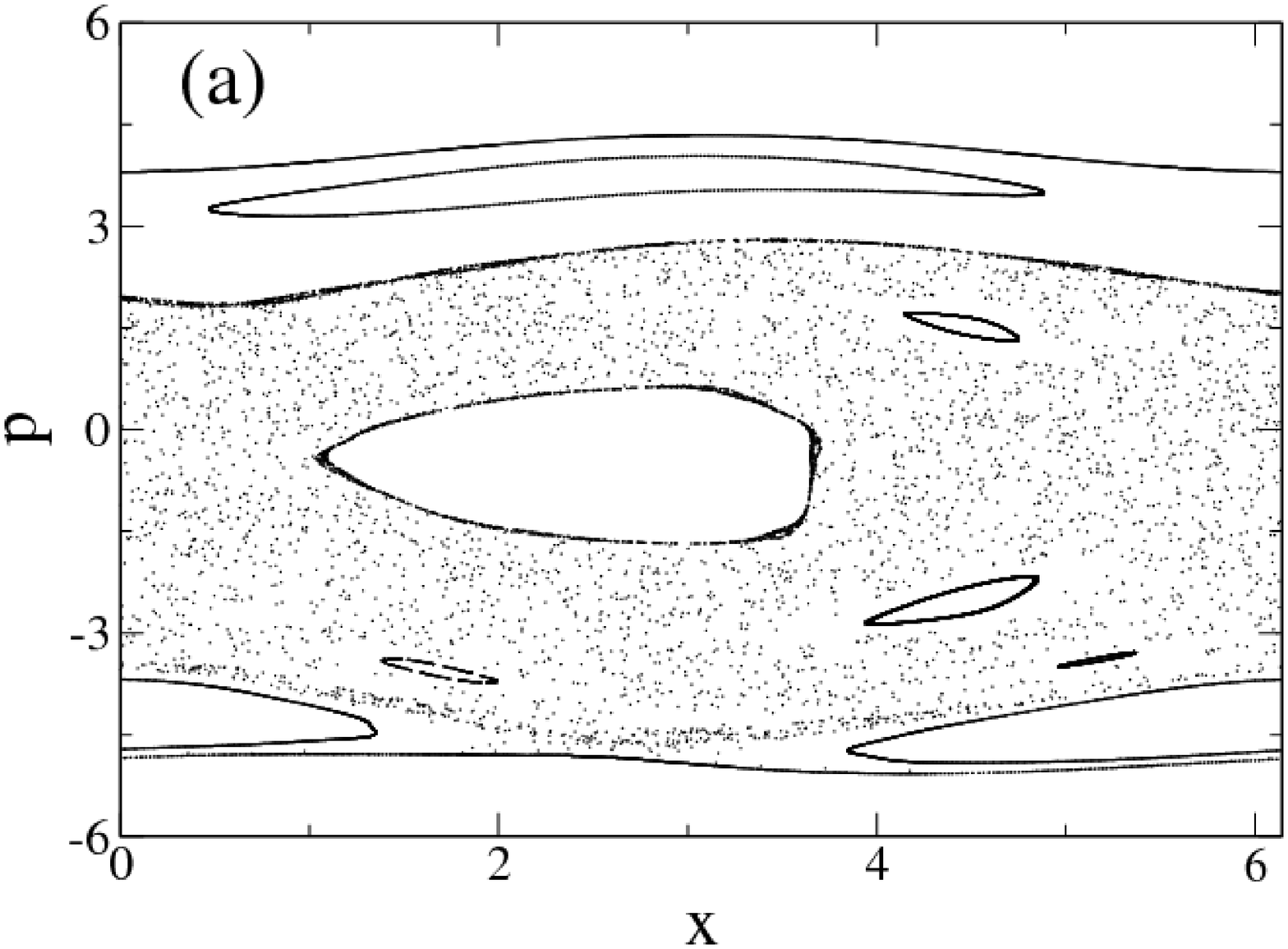}
\includegraphics[width=4.2cm,height=4cm]{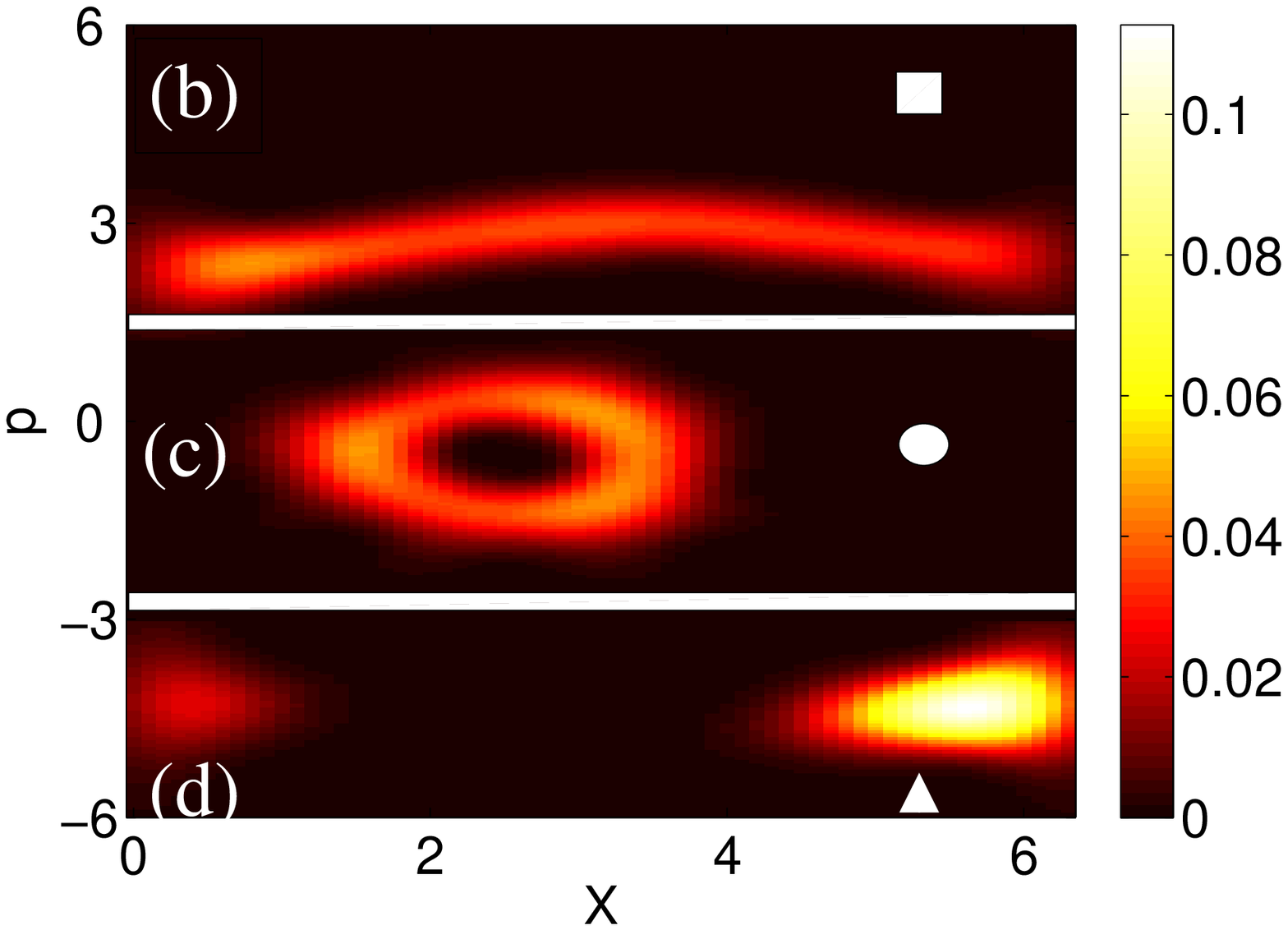}
\end{tabular}
\begin{tabular}{cc}
\includegraphics[width=4cm,height=4cm]{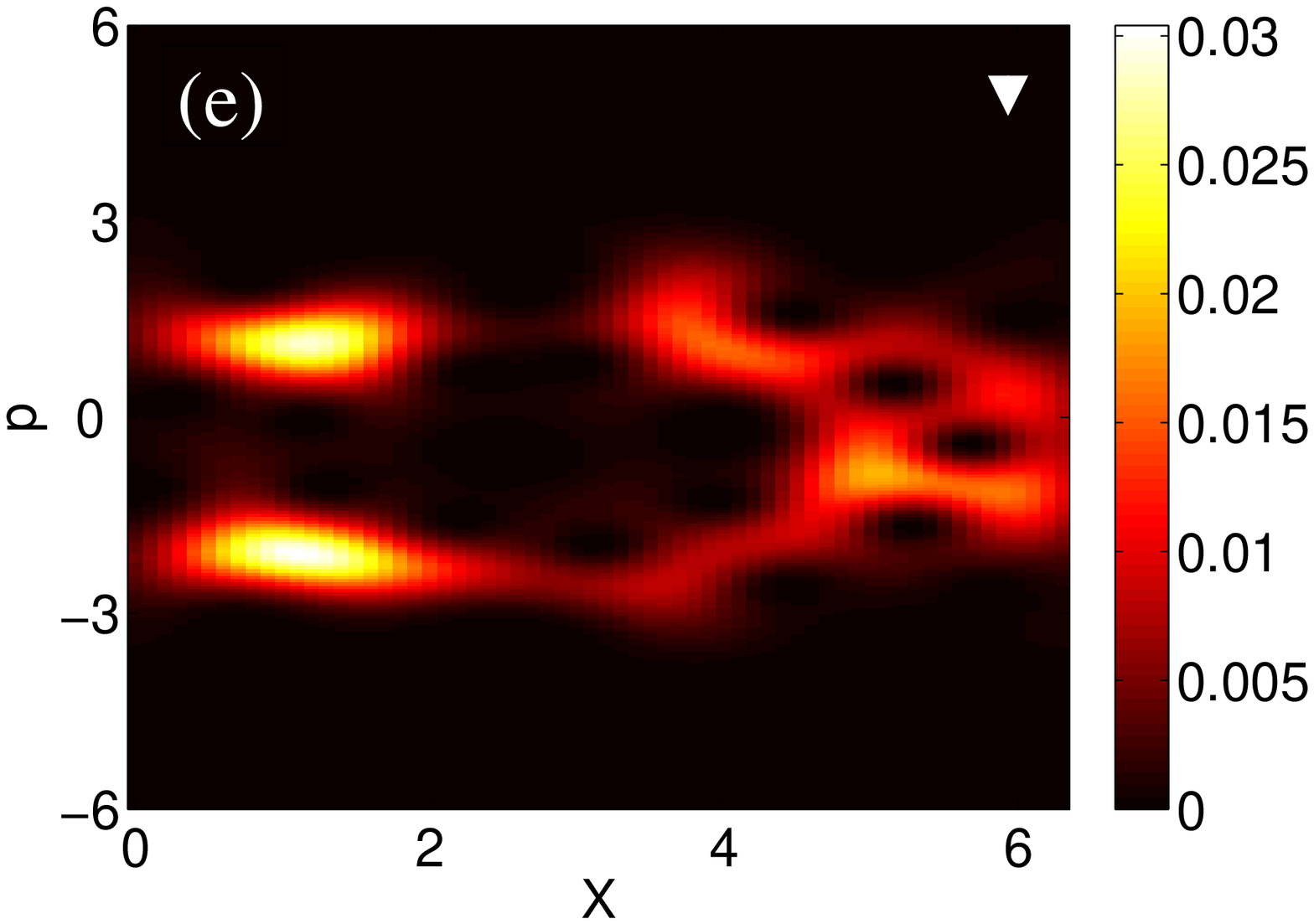}
\includegraphics[width=4cm,height=4cm]{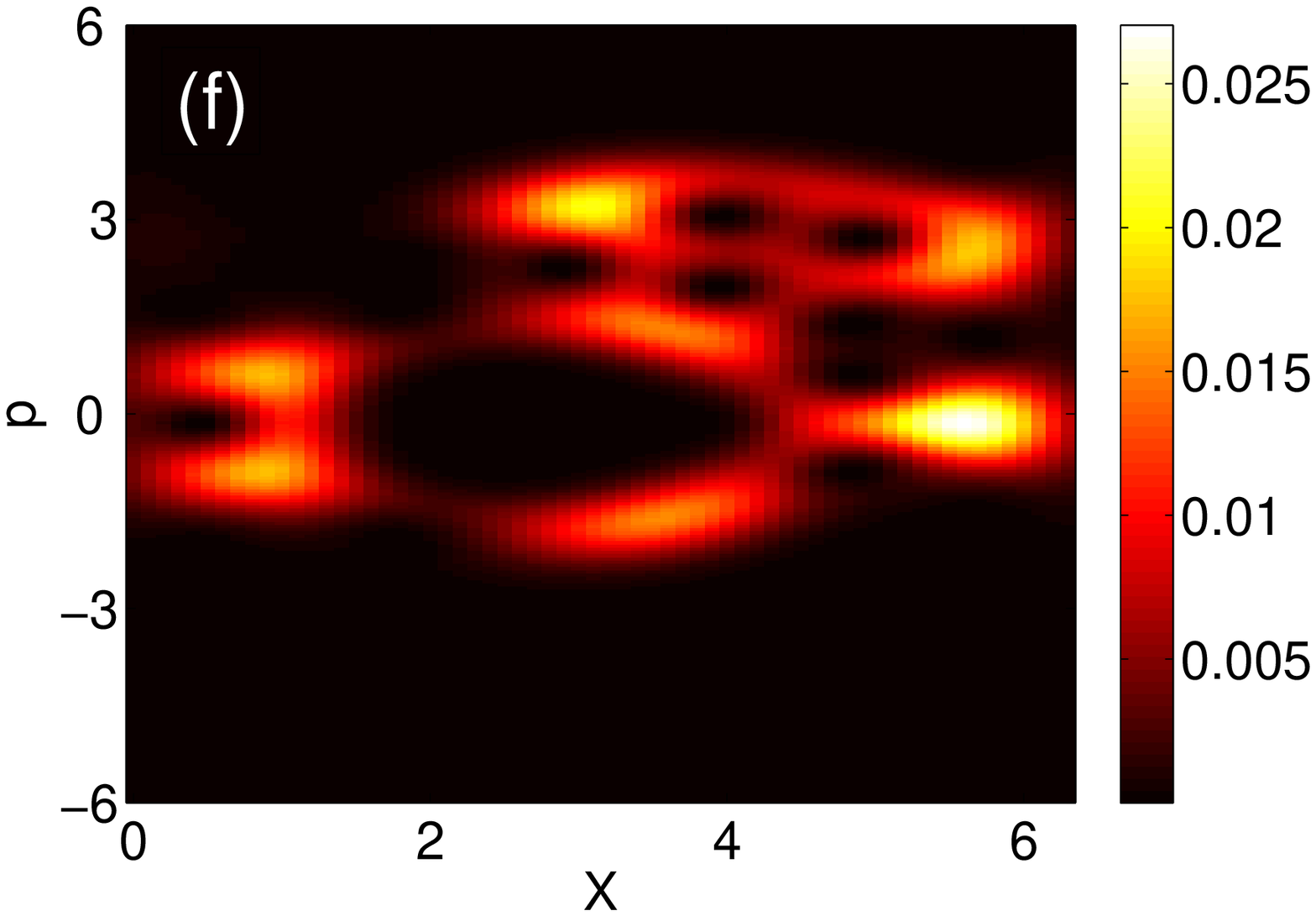}
\end{tabular}
\label{Figure1} \caption{(a) Poincar\'{e} section for the classical
limit; (b-f) Husimi representations for different Floquet eigenstates for
the Hamiltonian (\ref{eq:ham})  with $\hbar=0.2$ (momentum is in units of the
recoil momentum, $p_{r}=\hbar k_{L}$, with $k_{L}=1$). The
parameters are $E_{1}=E_{2}=2$, $\omega=2$, $\theta=-\pi/2$ and
$t_{0}=0$ for (b-e), and $E_{1}=3.26$, $E_{2}=1$,
$\omega=3$, $\theta=-\pi/2$ and $t_{0}=0$ for (f). } \label{Fig:husi}
\end{figure}
\end{center}

The Hamiltonian (\ref{eq:ham}) is periodic in time with period $T$.
The solutions  $|\psi_{\alpha}(t)\rangle =
U(t,t_{0})|\psi_{\alpha}(0)\rangle$ of the Schr\"{o}dinger equation
\begin{equation}
i \hbar \frac{\partial}{\partial t} |\psi(t)\rangle =
H(t,t_{0})|\psi(t)\rangle, \label{eq:Schrodin}
\end{equation}
can be characterized by the eigenfunctions of $U(t_0+T,t_0)$ which satisfy the Floquet theorem:
$ |\psi_{\alpha}(t)\rangle= e^{-i\frac{E_{\alpha}}{T}t}
|\phi_{\alpha}(t)\rangle$,
$|\phi_{\alpha}(t+T)\rangle=|\phi_{\alpha}(t)\rangle$. The
quasienergies $E_{\alpha}$ $(-\pi < E_{\alpha} < \pi)$ and the
Floquet eigenstates can be obtained as solutions of the eigenvalue
problem of the Floquet operator
\begin{equation}
U(T,t_{0})|\phi_{\alpha}(t_{0})\rangle = e^{-i
E_{\alpha}}|\phi_{\alpha}(t_{0})\rangle. \label{eq:Floquet}
\end{equation}
The Floquet eigenstates provide a complete orthonormal basis
and the stroboscopic quantum state can be
expressed as
\begin{equation}
|\psi(mT, t_{0})\rangle = \sum_{\alpha} C_{\alpha}(t_{0}) e^{-im
E_{\alpha}}|\phi_{\alpha}(t_{0})\rangle, \label{eq:expans}
\end{equation}
where the coefficients $\{C_{\alpha}\}$
depend on $t_0$. For later
convenience the integer $\alpha$, which sorts
the states $|\phi_{\alpha}\rangle$ such that the mean kinetic energy $\langle
p^{2}\rangle_{\alpha} \equiv 1/T \int_0^T \langle \phi_{\alpha} |
\hat{p}^2 | \phi_{\alpha} \rangle dt_0 $ monotonically increases.

With the help of a gauge transformation, $|\psi \rangle \rightarrow
\exp(-\frac{i}{\hbar}x\int_{0}^{t}E(t')dt') |\psi \rangle$
\cite{gauge1}, the solution of the time-dependent Schr\"{o}dinger
equation for the Hamiltonian (\ref{eq:ham}), may be written as
\begin{equation}
|\psi(t)\rangle =e^{-\frac{i}{\hbar} \int_{0}^{t} \{\frac{1}{2}[\hat{p}-
A(t',t_{0})]^{2}+(1+\cos x)\}dt'} |\psi(0)\rangle,
\label{eq:evolution}
\end{equation}
with the vector potential  $A(t,t_{0})=-\frac{E_{1}}{\omega}
\sin(\omega (t-t_{0}))-\frac{E_{2}}{2\omega} \sin(2 \omega
(t-t_{0})+\theta)$. Due to discrete translational invariance and
Bloch's theorem all Floquet states are characterized by
a quasimomentum $\kappa$ with
$|\phi_{\alpha}(x+2\pi)\rangle = {\rm e}^{i \hbar \kappa}
|\phi_{\alpha}(x)\rangle$.
\begin{figure}
\includegraphics[width=0.5\textwidth]{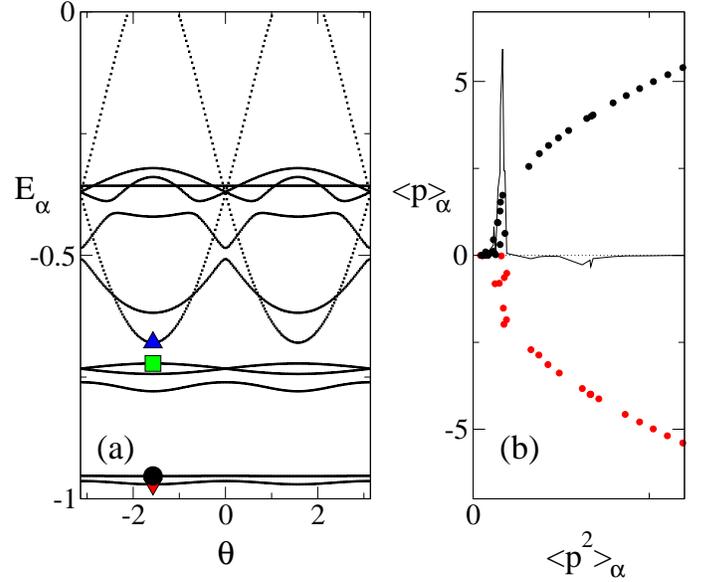}
\caption{(a) A part of the quasienergy  spectrum  as a function of
the parameter $\theta$. The symbols indicate the corresponding
Floquet states shown in Fig.\ref{Fig:husi}(b-e). (b) The mean
momentum $\langle p \rangle_{\alpha}$ vs mean kinetic energy
$\langle p^{2}\rangle_{\alpha}$ for the Floquet states (filled
circles). The line corresponds to the cumulative momentum,
$P_{\alpha}$ (see text). The parameters are the same as in
Fig.\ref{Fig:husi}(b-e). All the momenta are scaled in units of the
recoil momentum $p_{r}$.} \label{Fig:spectrum}
\end{figure}
Here we choose $\kappa=0$ which corresponds to initial states
where atoms equally populate all (or many) wells of the spatial
potential. The
wave function is expanded in the plane wave
eigenbasis of the momentum operator
$\hat{p}$, $|n \rangle=\frac{1}{\sqrt{2 \pi}} e^{i n x}$.
The Floquet propagator
$U(T,t_{0})$ is obtained by solving the Schr\"{o}dinger
equation over a single period $T$ for a
sufficiently large set of plane waves
$|n \rangle$ with $n=0, \pm 1, \pm 2, ..., \pm N$, where $2N+1$ is the
total number of basis states taken into account. The numerical calculations follow the
integration method described in Ref.\cite{gauge}, results are shown for $N=60$, which
do not depend upon further increase of $N$.

If the Hamiltonian is invariant under the shift symmetry $S_{a}$ (\ref{eq:Sa}),
then the Floquet matrix
has the property  $ U(T,t_{0})
=\left[U^{\maltese}\left(T/4,t_{0}\right)U\left(T/4,t_{0}\right)\right]^{T}U^{\maltese}
\left(T/4,t_{0}\right)U\left(T/4,t_{0}\right)$ \cite{Graham}.
Here $U^{\maltese}$ performs a transposition along the codiagonal of $U$.
In our case $S_a$ is always violated.

If the Hamiltonian is invariant under the time reversal symmetry $S_{b}$ (\ref{eq:Sb}),
then the Floquet matrix
has the property $U(T,t_{0})=U(T,t_{0})^{\maltese}$ \cite{Graham}.
That symmetry will be recovered for $\theta=0,\pi$. Then
the Floquet matrix has an irreducible representation using
even and odd basis states $|n\rangle_{s,a} = (|n\rangle \pm |-n\rangle)/\sqrt{2-\delta_{n,0}}$.

The mean momentum
expectation value  $J(t_0)= \lim_{t\rightarrow \infty} 1/t \int_{t_0}^{t}\langle
\psi(t,t_0)|\hat{p}|\psi(t,t_0)\rangle$
measures the asymptotic current.
Expanding the wave function over the Floquet states
the current becomes
\begin{equation}
J(t_0)=\sum_{\alpha} \langle p \rangle_{\alpha}
|C_{\alpha}(t_{0})|^{2}, \label{eq:current}
\end{equation}
where $\langle p \rangle_{\alpha}$ is the mean
momentum of the Floquet state $|\phi_{\alpha}\rangle$.

For the symmetric case $\theta=0,\pm \pi$ it follows that
$\langle p \rangle_{\alpha}=0$ for all $\alpha$. Consequently $J=0$ in this case.
We especially note that this is true for states with arbitrarily
large kinetic energy, for which the corresponding quasienergies become
almost pairwise degenerated.
For $\theta\neq 0,\pm \pi$ the Floquet states become
asymmetric. The quasidegeneracies are removed (Fig.\ref{Fig:spectrum}a).
Especially Floquet states with large kinetic energies
acquire large mean momenta (Fig.\ref{Fig:spectrum}b), thus becoming
transporting. This is a consequence of the fact that the true perturbation
parameter regulating the desymmetrization around the symmetric quasidegeneracy
points for small $\theta$ is proportional to $n E_1^2E_2 \theta$.
Fig.\ref{Fig:spectrum}b also shows the cumulative
average momentum, $P_{\alpha+2}=P_{\alpha}+\langle p
\rangle_{\alpha+1} +\langle p \rangle_{\alpha+2}$,  $P_{0}=\langle
p\rangle _0$. The asymmetry
stems mainly from Floquet states located in the chaotic layer
region of the classical phase space (see results below).
With increasing $\langle p^2 \rangle_{\alpha}$,
$P_{\alpha}$ goes to zero in full accordance with the
fact that total current over the whole momentum space should be zero
\cite{Ketz}.

In Fig.\ref{Fig:husi}(b-e) we present
Husimi distributions \cite{Husimi} for several Floquet states.
The dimensionless Planck constant $\hbar=0.2$ is in  a range, when it is
possible to establish a correspondence between different
Floquet states and the invariant manifolds of the mixed phase space
for the classical limit. Each plot carries a symbol, which shows the
corresponding location of the quasienergy of that state in Fig.\ref{Fig:spectrum}a.
The states (b-d) are located in various regular phase space regions,
while state (e) is located inside the chaotic layer.
\begin{figure}
\includegraphics[width=7cm,height=6cm]{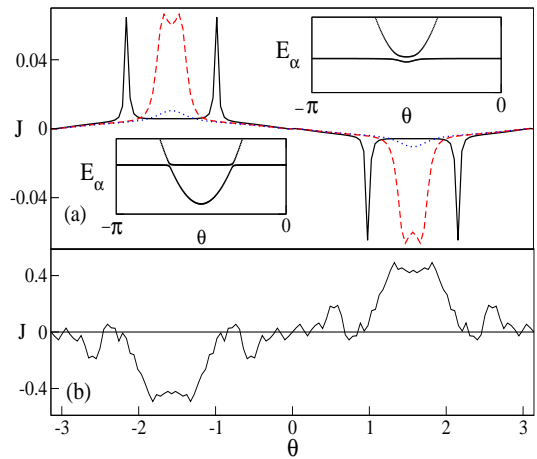}
\caption{ (a) The average current $J$ (in units of the recoil momentum)
vs $\theta$ for different amplitude values of the second harmonic,
$E_{2}$: $0.95$ (pointed line), $1$ (dashed line) and $1.2$ (solid
line). Insets: relevant details of the quasienergy spectrum
versus $\theta$ in the resonance region for
$E_{2}=1$ (top right) and $E_{2}=1.2$ (bottom left).
The parameters are $E_{1}=3.26$ and
$\omega=3$.
(b) The average current $J$ (in units of the recoil momentum)
vs $\theta$  for $E_{1}=3$, $E_{2}=1.5$ and
$\omega=1$.} \label{Fig:current}
\end{figure}

The current for an initial condition $|\psi(t_0)\rangle $ depends in
general on the initial phase $t_{0}$. Note that in the classical
case such an initial condition may also lead to some dependence of
the classical current on $t_0$, since the initial distribution may
overlap with different regular transporting manifolds. However, if
we start with a cloud of particles exactly located inside the
chaotic layer, the asymptotic current will be independent of $t_0$
for any choice of the distribution function over the chaotic
manifold. This is not true for the quantum case where the current
may even change its sign with the variation of $t_{0}$. It is a
consequence of the linear character of the Schr\"{o}dinger equation
\cite{Qchaos}. We will first discuss the results obtained after
averaging over the initial phase $t_{0}$. Then we can assign a
unique current value, $J=1/T \int_{0}^{T}J(t_{0})dt_{0}$, for fixed
parameters of the ac-field, $E_{1}$, $E_{2}$, and $\theta$.

Figure \ref{Fig:current}(a) shows the dependence of the average current
on the asymmetry parameter $\theta$ for the initial
condition $|\psi\rangle =|0\rangle $. The average current $J$ shows the
expected
symmetry properties  $J(\theta) = -J(\theta+\pi) = -J(-\theta)$.
On top of the smooth curves
we find several resonant peaks for $E_2=0.95$
where the current value changes drastically.
Comparing with the quasienergy spectrum, these
resonances can be unambiguously associated with avoided crossings
between two Floquet eigenstates. The Husimi distributions
show that one state locates in the chaotic layer,
and another one in a transporting island. Off resonance
the initial state mainly overlaps with the chaotic state,
which  yields some nonzero current. In resonance
Floquet states mix, and thus the new eigenstates contain
contributions both from the original chaotic state as well as from
the regular transporting island state. The Husimi distribution
of the mixed state is shown in Fig.\ref{Fig:husi}f, the strong asymmetry
is clearly observed. The regular island state has
a much larger current contribution, resulting in a strong enhancement
of the current.

>From an experimental point of view a too narrow resonance may become
undetectable due to resolution limitations. We thus studied how to
vary the width of the resonance without much affecting its
amplitude. It turns out to be possible by tuning another control
parameter, e.g. the amplitude $E_2$. We increase this field
amplitude in order to disentangle the two Floquet states and remove
the avoided crossing. That will happen for some value of $E_2$ at
$\theta=\pm \pi/2$. The details of the quasienergy spectrum around
that critical point are shown in the insets in
Fig.\ref{Fig:current}a. The two quasienergy spectra disentangle for
$E_2=1$ but stay close over a sufficient broader range of $\theta$
values. Thus the resonances become broader, as seen in
Fig.\ref{Fig:current}a. Further increase of $E_2$ to a value of 1.2
leads to a strong separation of the two spectra, and consequently to
a fast decay of the amplitude of the resonance.

The above numerical results for $\omega=2$ show a maximum current
value in the resonance region of the order of $0.06$ in units
of recoil momenta. In order to increase that output, we drive the system
into strongest resonance by choosing $\omega=1$, since that driver
frequency matches the oscillation frequencies of particles
at the bottom of the spatial periodic potential.
The result is shown in Fig.\ref{Fig:current}b. We again observe
a clear broad resonance, but the maximum current value increases
by an order of magnitude up to $0.5$ in units of recoil momenta.

As already mentioned, without averaging over $t_0$, the current
depends on the initial phase. However the observed resonance
structures are due to resonant interaction between Floquet states,
or avoided crossings of quasienergies. These resonances are
independent of the initial phase $t_0$. Indeed, in
Fig.\ref{Fig:temp} we plot the nonaveraged current as a function of
both $\theta$ and $t_0$. While the smooth background is barely
resolvable with the naked eye, the resonances are clearly seen, and
their position is not depending on $t_0$, while their amplitude
does. That implies that one can further maximize the resonant
current by choosing proper initial phases $t_0$, reaching values
above the recoil momentum.

Note that our approach is very different
from a recently proposed modified kicked rotor model \cite{kick}, where a biased
{\em acceleration} appears for an initial condition with preassigned nonzero
velocity. Here we study the regime of {\em stationary} current for the general case
of an
initial state of zero momentum for a model which has a finite current also in its classical counterpart
\cite{Den,D&F}.
\begin{figure}
\includegraphics[width=7cm,height=6cm]{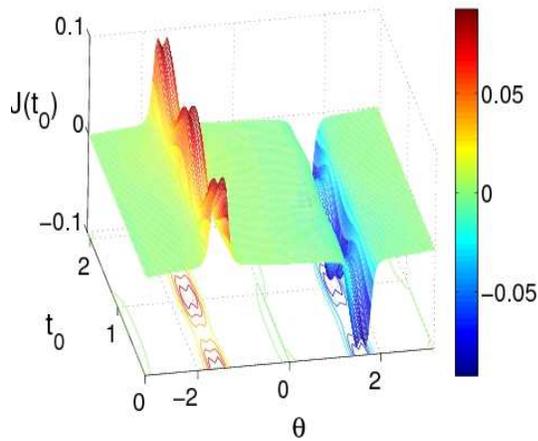}
\caption{Current dependence on the intial phase $t_0$ and $\theta$. The
 parameters are the same as in Fig.\ref{Fig:husi}f.}
\label{Fig:temp}
\end{figure}

In summary, we have studied the mechanisms of average current
appearance in driven quantum systems with broken
symmetries. The key source of such directed transport is the
desymmetrization of Floquet states. A peculiarity of the
quantum ratchet is the strong dependence of the current on the initial phase
of the applied field. Moreover, we found quantum
resonances induced by avoided crossings between Floquet states which
enhance the current drastically. Optimizing the drive frequency, amplitude
and initial phase, resonant currents easily reach the recoil momentum value and
should be experimentally observable using driven cold atoms in optical lattices.

We acknowledge fruitful discussions with F. Renzoni and M. Weitz.


\begin{thebibliography}{1000}

\bibitem{first} M. O. Magnasco, Phys. Rev. Lett. \textbf{71}, 1477
(1993); P. H\"{a}nggi and R. Bartusek, in \textit{Nonlinear Physics
of Complex Systems - Current Status and Future Trends}, edited by J.
Parisi, S. C. M\"{u}ller, and W. Zimmermann,  Lect. Notes. Phys.
\textbf{476} (Springer-Verlag, Heideberg, 1996), p. 294; F.
J\"{u}licher, A. Ajdari, and J. Prost, Rev. Mod. Phys. \textbf{69},
1269 (1997).

\bibitem{Reim} P. Reimann, Phys. Rep. \textbf{361}, 57 (2002).

\bibitem{Qu1} P. Reimann, M. Grifoni, and P. H\"{a}nggi, Phys. Rev.
Lett. \textbf{79}, 10 (1997); J. Lehmann et al,
ibid. \textbf{88}, 228305 (2002); M.
Grifoni et al,
ibid. \textbf{89}, 146801 (2002).

\bibitem{Flach1} S.~Flach, O.~Yevtushenko, and Y.~Zolotaryuk, Phys. Rev. Lett.
\textbf{84} 2358 (2000).

\bibitem{Flach2} O. Yevtushenko, S. Flach, Y. Zolotaryuk, and A. A. Ovchinnikov,
Europhys. Lett. \textbf{54}, 141 (2001).

\bibitem{Den} S. Denisov \textit{et al},  Phys. Rev. E
\textbf{66}, 041104 (2002).

\bibitem{Ketz} H. Schanz, M.-F. Otto, R. Ketzmerick, and T. Dittrich,
Phys. Rev. Lett. \textbf{87}, 070601 (2001); H. Schanz, T. Dittrich,
and R. Ketzmerick,  Phys. Review E \textbf{71}, 026228 (2005).

\bibitem{ren1} M. Schiavoni et al,
Phys. Rev. Lett. {\bf 90}, 094101 (2003);
P. H. Jones, M. Goonasekera, and F. Renzoni,
ibid. {\bf 93}, 073904 (2004); R. Gommers, S.
Bergamini, and F. Renzoni, ibid. {\bf 95}, 073003 (2005);
R. Gommers, S. Denisov, and F. Renzoni, ibid.
\textbf{96}, 240604 (2006).

\bibitem{Sis} J. Dalibard and C. Cohen-Tannoudji, J. Opt. Soc. Am. B
\textbf{6}, 2023 (1989); P. J. Ungar,
\textit{et al}., ibid. \textbf{6}, 2058 (1989).

\bibitem{rew} L. Guidoni and P. Verkerk,
Journal of Optics B {\bf 1}, R23 (1999); W. K. Hensinger et al,
ibid. {\bf 5}, R83 (2003).

\bibitem{Qu2}T. S. Monteiro et al,
Phys. Rev. Lett.
\textbf{89}, 194102 (2002); G. G. Carlo et al,
ibid. \textbf{94}, 164101 (2005).

\bibitem{D&F} S. Denisov and S. Flach, Phys. Rev. E \textbf{64}, 056236
(2001).

\bibitem{gauge1} When calculating Husimi distributions, average
kinetic energies, etc., we  used the inverse gauge transformation, in
order to return to the original wave function.

\bibitem{gauge} M. Latka, P. Grigolini, and B.J. West, \ Phys.\ Rev. \ A
\textbf{50}, 1071 (1994).

\bibitem{Graham} R. Graham and J. Keymer, Phys. Rev. A \textbf{44}, 6281 (1991).

\bibitem{Husimi} S.-J. Chang and K.-J. Shi, Phys. Rev. A \textbf{34}, 7
(1986).

\bibitem{Qchaos} F. Haake, \textit{Quantum signature of chaos}
(Springer-Verlag, London, 1991).

\bibitem{kick} E. Lundl and M. Wallin, \ Phys. \ Rev. \ Lett. \textbf{94}, 110603 (2005).
























\end{thebibliography}
\end{document}